# Detecting External Electron Spins Using Nitrogen-Vacancy Centers

H.J. Mamin, M.H. Sherwood, and D. Rugar

*IBM Research Division*
*Almaden Research Center*
*650 Harry Rd.*
*San Jose CA 95120 USA*

September 14, 2012

Near-surface nitrogen-vacancy (NV) centers have been created in diamond through low energy implantation of $^{15}$N to sense electron spins that are external to the diamond. By performing double resonance experiments, we have verified the presence of g=2 spins on a diamond crystal that was subjected to various surface treatments, including coating with a polymer film containing the free radical 2,2-diphenyl-1-picrylhydrazyl (DPPH). Subsequent acid cleaning eliminated the spin signal without otherwise disrupting the NV center, providing strong evidence that the spins were at the surface. A clear correlation was observed between the size of the detected spin signal and the relaxation time $T_2$ for the six NV centers studied. We have developed a model that takes into account the finite correlation time of the fluctuating magnetic fields generated by the external spins, and used it to infer the signal strength and correlation time of the magnetic fields from these spins. This model also highlights the sensitivity advantage of active manipulation of the longitudinal spin component via double resonance over passive detection schemes that measure the transverse component of spin.

*H. Jonathon Mamin (corresponding author)
mamin@almaden.ibm.com
408-927-2502

# I. INTRODUCTION

Nitrogen-vacancy centers are atomic-scale defects in diamond that have a host of attractive properties for nanoscale magnetometry.[1,2] They have a spin state (S=1) that is responsive to external magnetic fields, and long coherence times (>1ms in isotopically pure diamond) that enable nanotesla sensitivity at room temperature.[3] NV centers have been proposed as magnetic field transducers for nanoscale magnetic resonance imaging (nanoMRI),[4,5,6] making them a possible alternative to magnetic resonance force microscopy, which has comparable field resolution, and has produced 3D images of tobacco mosaic virus with better than 10 nm spatial resolution.[7]

NanoMRI requires sensing the fluctuating magnetic fields ("spin noise")[8,9] from ensembles of magnetic nuclei such as the protons within the sample of interest. Recently, detection of relatively distant $^{13}$C nuclei within the diamond lattice has been achieved using dynamic decoupling techniques, demonstrating the feasibility of detecting single nuclear spins as far away as ~3 nm, representing an impressive first step toward sensing more distant nuclei.[10,11] Still, if NV centers are to be useful transducers for more general purpose imaging, they must be able to detect nuclear spins that are external to the diamond. This requires having a center close to the surface of the diamond, so that it can be brought near the sample.

While nuclear spin detection remains one of the key motivations for NV magnetometry, detection of external electron spins is also an area of great interest, both as a proof of principle for detecting external spins, and perhaps as a way to indirectly detect nuclear spins using organic radicals[12] or other paramagnetic centers.[13] Using double electron-electron resonance (DEER), Grotz *et al.* have previously demonstrated detection of electron spin signals using surface implanted NV centers by treating the diamond with the spin-label TEMPO.[14] The signal had the characteristic frequency for a g=2 electron spin, and it disappeared when the molecule was removed. This suggests that the signal was due to the TEMPO, though no specific spectral fingerprint was seen which could uniquely identify the source. Because only a monolayer was used, the signals were probably arising from a small number of spins, possibly single spins. Dangling bonds on the diamond surface are another potential source of spin signal.

We were inspired by this work to take a similar approach, except that we substituted the monolayer of TEMPO with a relatively thick polymer film containing the molecule 2,2-diphenyl-1-picrylhydrazyl (DPPH). DPPH contains an electron free radical and is a common, well-characterized test sample for electron paramagnetic resonance. The thick film increases the number of spins in the detection volume, resulting in a net statistical spin polarization that should be easier to detect than a single spin. In this paper, we explore this statistical spin signal from a diamond surface coated with DPPH and subjected to various surface treatments. We also fit the results to a simple theoretical model that takes the finite correlation time of the electron spins into account.

## II. SAMPLE PREPARATION

The NV sample consisted of electronic grade diamond[15] that was implanted[16] with $^{15}N$ at 10 keV at a dose of $10^9$/cm$^2$. The relatively low energy insured that the NV centers were reasonably close to the surface (10-20 nm expected depth), and the use of $^{15}N$ allowed us to identify the implanted nitrogen though the hyperfine spectrum. After implantation, the samples were annealed at 900° C in a vacuum below $10^{-6}$ T for 1 hour, and cleaned in a boiling solution of sulfuric acid and potassium nitrate for 12 hours. Besides the initial acid cleaning, a variety of surface treatments were used. A solution of 5% DPPH by weight in polystyrene was prepared and dissolved in toluene at a concentration of 0.5% by weight. It was then spun onto the diamond substrate at 6000 rpm. The thickness of the film was non-uniform due to edge effects, with clear interference fringes visible at the edges. In the region of interest, the thickness is estimated optically to be below 100 nm. Subsequent measurements were made on the sample after (a) rinsing of the sample with toluene, (b) repeating the acid cleaning, and (c) re-applying the DPPH. In each case, we returned to the identical NV centers to compare the signals.

Fig. 1 shows the sample geometry, along with scanned confocal fluorescence images of a region containing a constellation of implanted NVs, both with the DPPH applied and after the second acid cleaning. The pattern of the NV centers in scanned fluorescent images appears unchanged by these treatments. Six NV centers were identified and studied in depth as candidates for double resonance experiments. All six showed $^{15}$N hyperfine splitting, indicating they were implanted centers, as opposed to naturally occurring $^{14}$N centers. The chosen NV centers were verified to be single centers based on the photon anti-bunching behavior. Three of the centers had short $T_2$ times as determined through spin echo measurements (roughly 20 µs or less), labeled 1-3, and three others, labeled 4-6, had relatively long $T_2$ times (greater than 100 µs).

### III. DOUBLE RESONANCE MEASUREMENTS

We performed double resonance on the NV centers using a variation of the protocol used in ref. 14. The protocol uses a Hahn spin echo to detect ac magnetic fields, where the fields are generated by one or more spins that are separate from the NV center (so-called "dark spins"). The dark spins are manipulated through application of a second microwave field at their Larmor frequency. We applied two identical π pulses to manipulate the dark spins (Fig. 2), one in each half of the echo, so that any inadvertent magnetic disturbance of the NV should be cancelled by the spin echo.[17]

When the rf frequency $f_D$ of the π pulses is at the Larmor frequency of the dark spins, the longitudinal magnetization of the spins will be inverted, creating a dipolar field that reverses in synchrony with the spin echo sequence applied to the NV center. The synchronous reversal of the dipolar field will act to disrupt the spin echo response. Figure 2(b) shows the NV optical signal as a function of the rf frequency of the π pulses applied to the dark spins, where the total echo time $\tau = 11.6$ μs. The external field was 320 G, putting the NV frequency $f_{NV} = 1.976$ GHz. A dip in the optical response is observed at $f_D = 0.896$ GHz. This corresponds to a g-factor $g = 2.00 \pm 0.01$ and is a clear indication that we are performing double resonance on a spin distinct from the NV center. Similar paramagnetic signals from surface-related spins have also been reported on both bulk[18,19,20] and nanodiamonds.[13]

In a second experiment, we used a single dark spin π pulse and varied the position $t$ of the pulse within the echo. Because the spin echo is sensitive to changes in the field between the first and second halves of the echo, flipping the dark spin in the center of the echo should result in the strongest signal. This is indeed the observed behavior as shown in Fig. 3, which plots the echo response as a function of the timing of the π pulse. This provided additional strong evidence that we are detecting dark spins via double resonance.

We can fit the observed $t$ dependence in Fig. 3 to a theoretical model that takes into account both the strength of the fluctuating field as well as the finite $T_1$ of the dark spin. Details of the model are given in Section V. From these fits, we infer $T_1$ greater than about 10 µs, and a rms fluctuating magnetic field strength of roughly 400 nT. To translate this field into number of dark spins, it is necessary to know the NV depth. We expect the depth to be roughly 10-20 nm based on the implantation energy (10 keV), but there can be significant variation due to straggle.[21] If we assume a mean separation of 20 nm between the NV center and the dark spins, this suggests that the signal arises from ~ 25 dark spins adding incoherently. However, we cannot rule out a signal arising from a single dark spin that is within about 10 nm from the NV center.

Changing the width $T$ of the pulse applied at the dark spin Larmor frequency should result in Rabi oscillations. Figure 4 shows the response as a function of $T$ with some rapidly decaying oscillations. The fact that only about one oscillation is observed indicates that the coherence time $T_2^*$ of the dark spin is comparable to the Rabi period, roughly 300 ns in this case.

## IV. SURFACE TREATMENT STUDIES

The above double resonance results provide convincing evidence that the NV center is able to detect some form of paramagnetic spins that are separate from the NV center, in agreement with previous reports.[14,18,19,20] The results do not, however, give unambiguous information as to the source of these spins. The spectrum in Fig. 2 is rather featureless, with little detectable fine structure. The observed width of roughly 20 MHz, (or ~ 7 G), is somewhat narrow compared to the expected width of roughly 12-20 G seen in similar samples of DPPH in polystyrene.[22] While DPPH does have a *g* value very close to 2.00,[23] so do many other paramagnetic centers. Thus, at our current level of precision, we cannot assign a unique identity to the dark spins based on the spectrum.

In an effort to elucidate the source of the dark spin or spins, we subjected our sample to a series of surface treatments. As demonstrated in Fig. 1, these treatments preserved the individual NV centers, allowing us to compare the dark spin signal before and after various surface treatments on the same centers. One such comparison is shown in Fig. 5 for NV1. The strongest signal was seen after initial application of the DPPH in polystyrene. The removal in toluene did not cause the signal to disappear, though it did become weaker. The fact that the toluene did not completely remove the signal suggests that the signal might not be due to the DPPH, though it is conceivable some residual monolayer remained.

On the other hand, another round of acid cleaning completely eliminated the signal to within the level of the noise, for both NV1 and NV2. This result confirms that these dark spins were surface-related. (Surprisingly, NV3 still showed a signal after the final cleaning step, suggesting that it may have been due to an internal or surface dark spin that was not destroyed by the acid cleaning.) A final test was to reapply the polystyrene/DPPH, which did not cause the signal to reappear, but rather gave the same results as post-acid cleaning: signal on NV3, but none on any of the others.

These results argue against DPPH being the origin of the dark spin signal, which raises the intriguing question as to why the DPPH would not be detected. It is possible that the $T_1$ of the DPPH is simply too short, which would reduce the detected signal as discussed in Section V. Nonetheless, we have strong evidence that different surface treatments can change the presence of surface-related electron spin centers.

An interesting correlation was observed between the presence or absence of a dark spin signal and the $T_2$ of the NV center. NV4, which had a rather long $T_2 \sim 100$ μs, did not show a dark spin signal, while NV2, which had a short $T_2$, ~10 μs, did show a signal (Fig. 6(a)). The correlation is evident in the plot in Fig. 6(b), which plots strength of signal vs $T_2$ under various conditions for all six NV centers. For the most part, the surface treatments did not significantly affect $T_2$. While we initially expected that the proximity of the dark spin might be the cause of the short $T_2$, we did not observe an increase in $T_2$ upon loss of the dark spin signal; if anything, the $T_2$ became slightly shorter. We speculate that the three NV centers that initially showed dark spin signal were all quite close to the surface, making them vulnerable to surface-induced decoherence.[24]

## V. MODELLING THE EFFECT OF FINITE CORRELATION TIMES

The use of NV centers to detect magnetic resonance from small ensembles of external spins will inevitably require the ability to measure stochastic signals arising from statistical polarization.[25,26] The frequency spectrum of these signals will be determined by the particular detection scheme used. For the active manipulations employed in the DEER scheme, the reversal frequency is chosen to match the NV spin echo time, which can result in kHz frequency signals. In other protocols, such as the Carr-Purcell-Meiboom-Gill (CPMG) protocol,[27] the frequency may be the natural Larmor frequency of the external spins, typically in the MHz range for nuclear spins and GHz range for electron spins. CPMG and related multipulse protocols have indeed been used successfully for detection of nuclear magnetic resonance (NMR) signals from internal nuclear spins (e.g. $^{13}$C) using NV centers.[10,11]

Whatever the method, the finite correlation time $T_C$ of the external signal can disrupt the synchronization of the signal with the NV detection protocol and thereby reduce the detected signal. For detection of the longitudinal spin signal, $T_C$ will be determined by the spin-lattice relaxation time $T_1$, while for transverse (Larmor frequency) detection, $T_2^*$ determines the correlation time. Often it is assumed that $T_C$ is much longer than the $T_2$ of the NV center, but this need not be the case.

Here we derive relatively simple analytic expressions for the NV response for arbitrary $T_C$, both for the DEER scheme as well as for signals at the Larmor frequency, as might be used for NMR detection. We first consider a simple Hahn echo, both with and without the active DEER manipulation. We assume that the dark spins give rise to an exponentially correlated Gaussian fluctuating field $B(t)$ with correlation time $T_1$, the longitudinal relaxation time. (The case of arbitrary $B(t)$ has also been analyzed by Hall *et al* using Taylor series expansions; this treatment did not focus on active manipulation of the dark spins.[28]) Following along the lines of Taylor *et al*,[29] we take the correlation function to be

$$\langle B(t)B(t')\rangle = B_{rms}^2 \exp(-|t-t'|/T_1), \tag{1}$$

where $B_{rms}$ characterizes the rms amplitude of the fluctuating field, and $\langle...\rangle$ denotes the average value. For the simple Hahn echo with no DEER pulse, the random phase $\delta\phi$ accumulated during any given echo is

$$\delta\phi = \gamma\int_0^{\tau/2} B(t)\, dt - \gamma\int_{\tau/2}^{\tau} B(t)\, dt, \tag{2}$$

where $\tau$ is the full spin-echo time. The spin echo response is then given by $S = S_0 + \Delta S \cos(\delta\phi)$ (assuming the phases of the microwave pulses are chosen appropriately). Here $S_0$ represents the average fluorescence of the two spin states and $\Delta S$ takes into account the contrast between the high fluorescence and low fluorescence states. For a randomly fluctuating field, the mean signal is given by[30]

$$S = S_0 + \Delta S \langle \cos(\delta\varphi) \rangle = S_0 + \Delta S \exp\left(-\langle (\delta\varphi)^2 \rangle / 2\right). \tag{3}$$

For small $\delta\phi$, this reduces to

$$\Delta S \langle \cos(\delta\phi) \rangle \; S = S_0 + \Delta S \left(1 - \langle (\delta\phi)^2 \rangle / 2\right).$$

The mean square phase accumulated during the echo evolution time is given by

$$\langle (\delta\phi)^2 \rangle = \left\langle \left[ \gamma\int_0^{\tau/2} B(t)\, dt - \gamma\int_{\tau/2}^{\tau} B(t)\, dt \right]^2 \right\rangle \tag{4}$$

$$= \left\langle \left[ \gamma\int_0^{\tau/2} B(t)\, dt - \gamma\int_{\tau/2}^{\tau} B(t)\, dt \right] \cdot \gamma\left[ \int_0^{\tau/2} B(t')\, dt' - \gamma\int_{\tau/2}^{\tau} B(t')\, dt' \right] \right\rangle$$

$$= \gamma^2 \left[ \int_0^{\tau/2}\int_0^{\tau/2} \langle B(t)B(t') \rangle\, dt\, dt' - 2\int_0^{\tau/2}\int_{\tau/2}^{\tau} \langle B(t)B(t') \rangle\, dt\, dt' + \int_{\tau/2}^{\tau}\int_{\tau/2}^{\tau} \langle B(t)B(t') \rangle\, dt\, dt' \right]$$

$$= \gamma^2 B_{rms}^2 \left[ \int_0^{\tau/2}\int_0^{\tau/2} e^{-|(t-t')|/T_1}\, dt\, dt' - 2\int_0^{\tau/2}\int_{\tau/2}^{\tau} e^{-|(t-t')|/T_1}\, dt\, dt' + \int_{\tau/2}^{\tau}\int_{\tau/2}^{\tau} e^{-|(t-t')|/T_1}\, dt\, dt' \right]$$

$$= \gamma^2 B_{rms}^2 \left[ 8T_1^2 e^{-\tau/2T_1} - 2T_1^2 e^{-\tau/T_1} - 6T_1^2 + 2T_1\tau \right].$$ In the limit of $T_1 \gg \tau$, this becomes $\gamma^2 B_{rms}^2 \tau^3 / 6T_1$, as found by Taylor.[29]

Next we consider the case where the active DEER pulse is added during the course of the Hahn echo. We make the basic assumption that the effect of a $\pi$ pulse applied at time $t_D$ to the external (dark) spins is to invert the sign of $B(t)$ for all subsequent times. If $t_D = \tau/2$ (middle of the Hahn echo), the expression for $\langle(\delta\phi)^2\rangle$ then becomes

$$\langle(\delta\phi)^2\rangle = \left\langle\left[\gamma\int_0^{\tau/2} B(t)\,dt + \gamma\int_{\tau/2}^{\tau} B(t)\,dt\right]^2\right\rangle, \tag{5}$$

where here $B(t)$ represents the field absent the inversion. Note that in the limit of very short coherence time, the cross-terms vanish in both Eq (4) and Eq (5), and the results are independent of whether or not a DEER pulse is applied, which is physically reasonable. In the limit of very long coherence time, the DEER pulses create a square wave time variation in the field synchronized to the echoes, and the response becomes maximal.

For arbitrary $T_1$ and $t_D$, Eq. (4) can be generalized by replacing $B(t)$ with $-B(t)$ where appropriate. After some algebra, we find

$$\langle(\delta\phi)^2\rangle = \gamma^2 B_{rms}^2 \tau^2\, f(\tau, t_D, T_C)\,, \text{ where}$$

$$f(\tau, t_D, T_1) \equiv 2\frac{T_1}{\tau^2}\left[\tau - 5T_1 + 4T_1 e^{-|\tau/2 - t_D|/T_1} - 2T_1 e^{-(\tau/2 + |\tau/2 - t_D|)/T_1} + 2T_1 e^{-(\tau/2 - |\tau/2 - t_D|)/T_1} + T_1 e^{-\tau/T_1}\right]. \tag{6}$$

For the special case where the DEER pulse is applied in the center of the echo,

$$f(\tau, t_D = \tau/2, T_1) = \frac{2T_1}{\tau^2}\left[\tau - T_1 + T_1 e^{-\tau/T_1}\right],$$ a dimensionless factor that approaches 1 monotonically for $T_1 \gg \tau$, as shown in Fig. 7(a). Thus, the effect of the finite correlation time is to reduce the influence of the dark spin field inversion on the spin echo by an amount that depends on the correlation time, which is again physically reasonable. Figure 7(a) also plots the same function for other values of the flip time $t_D$. Depending on where in the echo the pulse is applied, the function can show non-monotonic behavior as a function of $T_1$.

Although this analysis has been performed for spins whose field has a noise spectral density centered at DC, as is appropriate for $T_1$ processes, it is readily extended to spins precessing at the Larmor frequency, as might be sensed in a CPMG protocol, for example. In this case, the correlation function is determined by the $T_2^*$ coherence time and is taken to be $\langle B(t)^* B(t') \rangle = B_{rms}^2 \exp(-|t-t'|/T_2^*) \cos(\omega_0(t-t'))$. We can use this correlation function to calculate $\langle (\delta\phi)^2 \rangle$ in the case of a Hahn echo using Eq. (1), where we now choose the echo time $\tau$ to equal $2\pi/\omega_0$. The result can be written as

$$\langle (\delta\phi)^2 \rangle = \gamma^2 B_{rms}^2 \tau^2 \, h(\tau, T_2^*) \ , \text{where}$$

$$h(\tau, T_2^*) \equiv \frac{2T_2^*}{(\tau^2 + 4\pi^2 T_2^{*2})^2} \Big( 12\pi^2 T_2^{*2} - 3T_2^* \tau^2 + \tau^3 + 4\pi^2 T_2^{*2} \tau - T_2^* \tau^2 \exp(-\tau/T_2^*)$$
$$- 4T_2^* \tau^2 \exp(-\tau/2T_2^*) + 16\pi^2 T_2^{*3} \exp(-\tau/2T_2^*) \Big) \ .$$

## VI. NV SPIN ECHO SIGNAL

We can now use the calculated effects of finite dark spin correlation time to derive the expected measured NV spin echo response as a function of the various parameters. We first define $\alpha_0$ to be the number of photons counted per echo for the NV left in the bright ($m_s = 0$) state, and $\beta_0$ to be the counts per echo for the dark ($m_s = \pm 1$) state. For finite echo times, the apparent contrast between bright and dark states is reduced due to the finite coherence time $T_{2,NV}$ of the NV center. We explicitly account for this by writing the counts per echo for arbitrary echo time as

$$\alpha(\tau) = \frac{\alpha_0 + \beta_0}{2} + \left( \frac{\alpha_0 - \beta_0}{2} \right) \exp\left[ -\left( \tau/T_{2,NV} \right)^n \right] \text{ and}$$

$$\beta(\tau) = \frac{\alpha_0 + \beta_0}{2} - \left(\frac{\alpha_0 - \beta_0}{2}\right) \exp\left[-\left(\tau/T_{2,NV}\right)^n\right],$$

where the exponent $n$ depends on the source of the NV decoherence.[31] These functional forms are chosen so that for long echo time $\tau$, $\alpha(\tau)$ and $\beta(\tau)$ both approach the mean count rate $(\alpha_0 + \beta_0)/2$, and for arbitrary time $\tau$, the difference in counts per echo shows the expected decay:

$$\alpha(\tau) - \beta(\tau) = (\alpha_0 - \beta_0) \exp\left[-\left(\tau/T_{2,NV}\right)^n\right].$$

We can combine Eq. (6) and Eq. (3) with the above definitions to derive the expected measured NV spin echo response as a function of the various parameters. We find that the fluorescence signal (normalized to $\alpha_0$) is

$$S = \frac{\alpha(\tau) + \beta(\tau)}{2\alpha_0} + \frac{\alpha(\tau) - \beta(\tau)}{2\alpha_0} \exp\left(-\gamma^2 B_{rms}^2 \tau^2 \, \mathrm{f}(\tau, t_D, T_1)/2\right). \qquad (7)$$

We have fit the measured DEER data to Eq. (7) for various choices of parameters, as shown in Fig. 3. The fitting parameters are the correlation time $T_1$ and the fluctuating field strength $B_{rms}$. (We independently determined $\alpha(\tau)/\alpha_0 = 0.91$ and $\beta(\tau)/\alpha_0 = 0.70$ in a separate experiment where the NV center at the end of the echo was left either in the high or low fluorescence state.) The effect of shorter $T_1$ can be largely offset by increased $B_{rms}$, so the fits are not particularly unique. For short $T_1$ times, however, there is a pronounced hook at small times $t_D$, which may be visible in the data, but it is not clear.[32]

For $T_1$ from 10 μs to essentially infinity, reasonable fits to the data could be obtained, with rms field strengths of roughly 400 nT-rms. As $T_1$ becomes short compared to $\tau = 17.2$ μs (the echo time), the quality of the fit rapidly declines, with a very poor fit at $T_1 = 2$ μs (Fig. 3(b) dashed green curve). The expected $T_1$ for DPPH under roughly similar conditions[22] is about 6 μs, which gives a fit that is still somewhat marginal (dashed blue curve). Given the overall uncertainties in both the fit parameters and the literature values, the role of DPPH here remains an open question.

In the case of Larmor detection, we can write an analogous expression for the signal

$$S = \frac{\alpha(\tau)+\beta(\tau)}{2\alpha_0} + \frac{\alpha(\tau)-\beta(\tau)}{2\alpha_0}\exp\left(-\gamma^2 B_{rms}^2 \tau^2\, \mathrm{h}(\tau,T_2^*)/2\right).$$

In principle it is possible to use these expressions to solve for the optimum echo time $\tau$ as a function of $T_{2,NV}$ and the correlation time, but we have not found closed-form solutions.

## VII. MINIMUM DETECTABLE MAGNETIC FIELD FLUCTUATIONS

Based on the usual assumption that the noise is determined by the shot noise in the photon counts,[29] we can use the above expressions to determine a minimum detectable mean-square field $B_{min}^2$ (i.e. one which results in unity signal-to-noise ratio). For small signals ($\gamma B_{min} \tau < \pi/4$), and assuming a dark spin correlation time $T_C$, acquisition time $T_a = N\tau$, and shot noise $\sqrt{N\alpha(\tau)}$, where $N$ is the number of echoes, we find

$$B_{min}^2 = \frac{4\sqrt{\alpha(\tau)}}{\gamma^2 \tau^{3/2}\left(\alpha(\tau)-\beta(\tau)\right)\sqrt{T_a}}\frac{1}{\mathrm{f}(\tau,\tau/2,T_C)} \quad (8a)$$

for active manipulation and

$$B_{min}^2 = \frac{4\sqrt{\alpha(\tau)}}{\gamma^2 \tau^{3/2}\left(\alpha(\tau)-\beta(\tau)\right)\sqrt{T_a}}\frac{1}{\mathrm{h}(\tau,T_C)} \quad (8b)$$

for Larmor detection.

Figure 7(b) plots the results as a function of correlation time $T_C$, assuming $\alpha(\tau) = 0.05$ counts per echo, $\beta_0/\alpha_0 = 0.7$, $\tau = T_{2,NV} = 100\ \mu s$, and $T_a = 1$ s. (Here we do not consider possible improvements in the effective $T_{2,NV}$ from using dynamic decoupling protocols such as CPMG.[29,33]). The dotted blue line is for active manipulation, and the solid line is for Larmor detection. Under these conditions, the curves are quite similar, except for saturating at different values for long $T_C$.[34] A more significant difference is that for the active longitudinal manipulation, the relevant correlation time $T_C = T_1$, which is typically much longer than $T_2^*$, the relevant correlation time for passive Larmor frequency detection. For example, if $T_1 = 1$ ms (blue dot) and $T_2 = 10\ \mu s$ (black square), the minimum detectable mean-square field is nearly 50 times less (7× lower in amplitude) with active manipulation of the dark spins compared to Larmor detection.

## VIII. CONCLUSION

Double resonance experiments were performed to detect paramagnetic centers (dark spins) on the surface of diamond via their influence on near-surface NV centers. By subjecting the surface to acid cleaning, we were able to remove the paramagnetic centers without otherwise disrupting the NV center, demonstrating that the dark spins were external to the diamond lattice. A correlation was observed between dark spin signal and NV coherence time, where only those NV centers with short $T_2$ detected dark spin signals. Even though the exact nature of the dark spins remains unclear, these observations have implications for the future detection of nuclear spins external to the diamond, where the ability to prepare spin-free surfaces will undoubtedly be important. Finally, we find that active manipulation of longitudinal magnetization, where the relevant correlation time is of order $T_1$ of the dark spins, is expected to be much more sensitive than passive schemes that detect signals at the Larmor frequency.

**Acknowledgments**

We gratefully acknowledge support from the DARPA QUASAR program. We also thank D. Awschalom, J. Wrachtrup, A. Yacoby, M. Lukin, R. Walsworth, P. Hemmer, A.C. Bleszynski Jayich, D. Twitchen, G. Balasubramanian, F. Reinhard, P. Maletinsky, M. Grinolds and M. Kim for helpful discussions.

[30] For a normally distributed variable *x* with zero mean and variance $\sigma^2$, the mean of $\cos(x)$ equals $\int_{-\infty}^{\infty} \frac{1}{\sqrt{2\pi\sigma^2}} e^{-x^2/(2\sigma^2)} \cos(x) dx = e^{-\sigma^2/2}$.

[31] L. Childress, M. V. Gurudev Dutt, J. M. Taylor, A. S. Zibrov, F. Jelezko, J. Wrachtrup, P. R. Hemmer, M. D. Lukin, Science **314**, 281 (2006).

[32] The hooks are related to the non-monotonic behavior seen in Fig. 7(a). This behavior can occur when the correlation time is of the order of the echo time. In these cases, the natural fluctuations are well-matched to the echo, giving some disruption of the echo pulse even without a DEER pulse. Then, for particular times, the DEER pulse acts to "undo" the disruption. Monte Carlo simulations of a bath of fluctuating spins show qualitatively the same behavior.

[33] R. de Sousa, Top. Appl. Phys. 115, 183 (2009).

[34] The saturation values differ by roughly a factor of 0.35, which is close to the expected factor of $(2/\pi)^2 = 0.40$ arising from the difference in phase accumulated from a square wave signal (DEER protocol) and a sinusoidal signal (Larmor precession).

**Figures and Figure Captions**

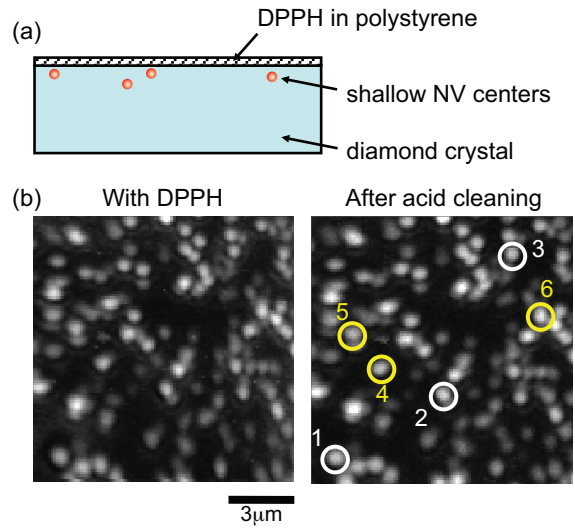

Fig. 1. (a) Sample geometry for detection of paramagnetic centers using NV centers in diamond. The NV centers are created close to the surface through ion implantation of $^{15}$N at 10 keV. A film of DPPH in polystyrene is then spun onto the diamond substrate to provide the sample material. (b) Fluorescent imaging showing the identical pattern of NV centers taken before (left) and after (right) removal of the DPPH film by acid cleaning. The specific NV centers used in the measurements are circled. Double resonance signals were observed on 1, 2 and 3, but not on 4, 5 and 6.

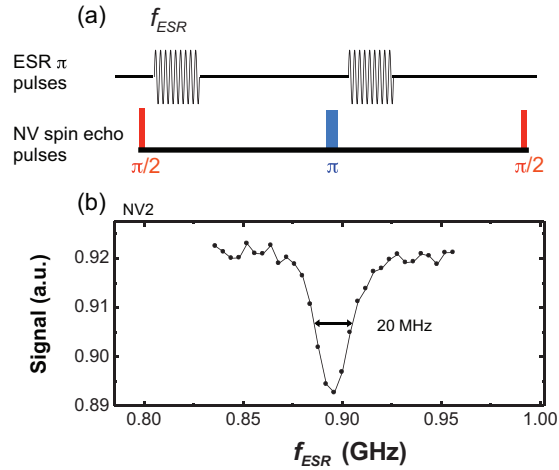

Fig. 2. (a) Protocol for performing double resonance to detect electron spins ("dark spins") in proximity to a NV center. $\pi$ pulses are applied at the Larmor frequency of the g=2 external spins, while performing a Hahn echo on the NV centers. Any change in magnetic field at the NV center caused by the reversal of the dark spin will disrupt the spin echo. (b) Optical fluorescence from NV2 as a function of the microwave frequency applied to the dark spin. A clear response is observed at 896 MHz, corresponding to g=2. The total echo time was 11.6 μs.

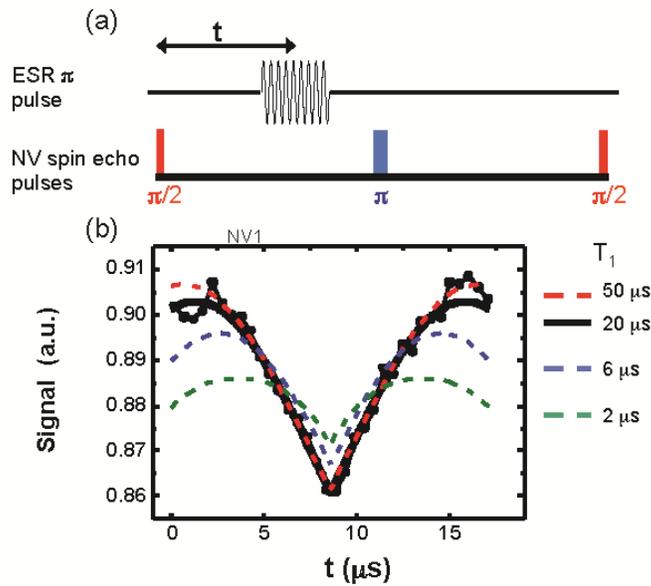

Fig. 3. Effect of pulse position within the Hahn echo. (a) Protocol uses fixed pulse width and frequency while varying the pulse position $t$. (b) NV echo signal *vs* pulse position $t$. Data are represented by solid circles. The lines are calculated using a 2-parameter fit to a model that assumes a classical ensemble of randomly fluctuating spins. $(T_1, B_{rms}) = (50\ \mu s,\ 370\ \text{nT-rms})$, (dashed red line); $(20\ \mu s,\ 425\ \text{nT-rms})$ (solid black); $(6\ \mu s,\ 500\ \text{nT-rms})$ (dashed blue); $(2\ \mu s,\ 700\ \text{nT-rms})$ (dashed green). Excellent fits to the data are obtained for $T_1 \geq 20$ μs and a signal strength of roughly 400 nT-rms, The total echo time was 17.2 μs. (color online)

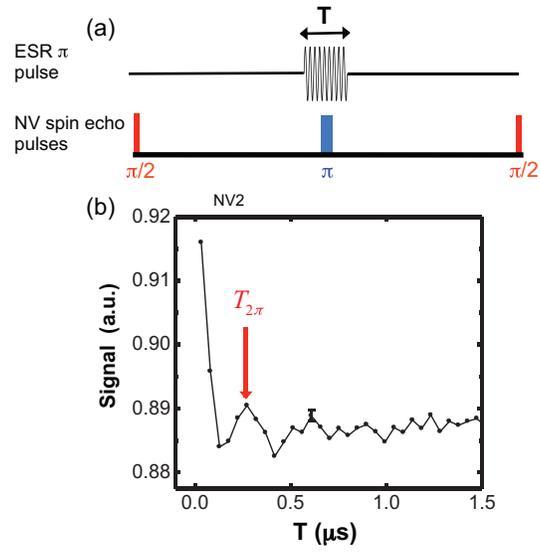

Fig. 4. Effect of pulse width on the Hahn echo. (a) Pulse position is fixed within the center of the echo, while the width $T$ is varied. (b) NV echo signal vs $T$, showing weak Rabi oscillations of the dark spin.

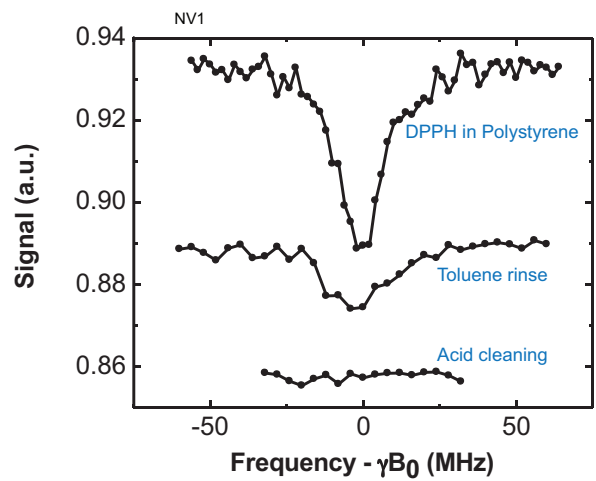

Fig. 5. Surface treatment studies of the double resonance signal for NV1. Uppermost curve was taken with the DPPH film present, middle curve is after a rinse in toluene, and lower curve is after acid cleaning the diamond (curves offset for clarity). Acid cleaning resulted in a total loss of signal (within measurement error). NV2 showed very similar results, while for NV3, some signal remained even after acid cleaning.

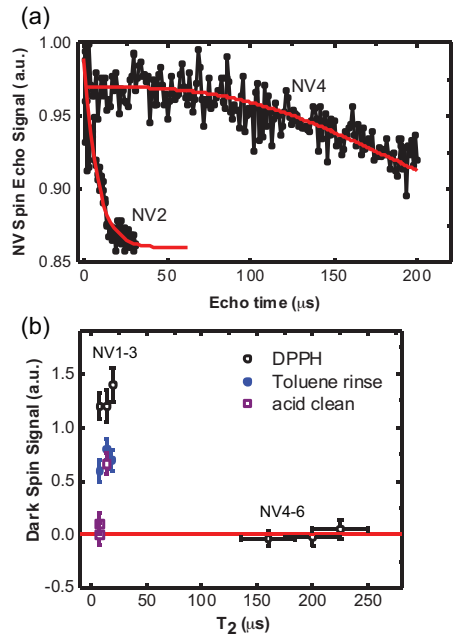

Fig. 6. (a) Spin echo signal *vs* echo time (no DEER pulse) for NV2 that detects the dark spins and NV4 that does not. (b) Correlation between dark spin signal and the $T_2$ of the NV center. The NV centers with the dark spin signal showed consistently lower $T_2$ times than the centers with no detectable signal. It is known that the ion implantation produces a distribution of depths, and one might expect deeper centers to be less strongly coupled both to the dark spins and the environment, resulting in longer $T_2$ times.

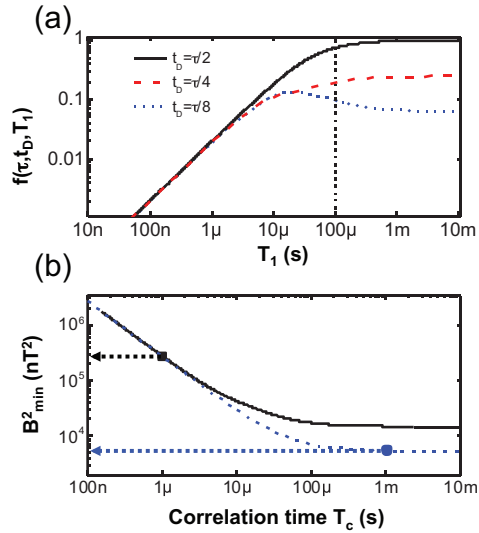

Fig. 7. Calculated effect of finite dark spin correlation times on detected signal. (a) $f(\tau,t_D,T_1)$ as a function of correlation time $T_1$ for DEER pulse applied at three values of $t_D$ for $\tau = 100$ μs (dotted vertical line). The function $f(\tau,t_D,T_1)$ represents a derating factor for the detected mean square field. When the pulse is applied in the middle of the echo ($t_D = \tau/2$), $f(\tau,t_D,T_1)$ is a monotonic function of $T_1$ that approaches 1 for large $T_1$. (b) Calculated minimum detectable mean square magnetic field *vs* field correlation time $T_C$ for two detection schemes: a longitudinal manipulation protocol (dashed blue curve) and a passive transverse detection scheme that matches the spin echo to the natural precession frequency of the dark spins (solid black curve). $\tau = 100$ μs for both cases. The correlation time $T_C$ depends on the protocol. Since $T_1$ is generally much longer than, for example, $T_2^*$ (blue dot *vs* black square), the longitudinal manipulation protocol will be considerably more sensitive than passive transverse detection schemes. (color online)